\documentclass[12pt]{iopart}
\usepackage{graphicx}

\begin{document}
\def \ee {\varepsilon}
\thispagestyle{empty}
\title[Fast Track Communication]{
Pulsating Casimir force
}

\author{
G.~L.~Klimchitskaya,${}^{1,2}$
U.~Mohideen,${}^3$
and V.~M.~Mostepanenko,${}^{1,4}$
}

\address{${}^1$Center of Theoretical Studies and Institute for Theoretical 
Physics, Leipzig University,
Postfach 100920, D-04009, Leipzig, Germany}

\address{$^2$North-West Technical University, Millionnaya St. 5,
St.Petersburg, 191065, Russia
}
\address{$^3$
Department of Physics and Astronomy, University of California, Riverside,
CA 92521, USA}

\address{$^4$
Noncommercial Partnership  ``Scientific Instruments'', 
Tverskaya St. 11, Moscow, 103905, Russia.}

\begin{abstract}
Based on the Lifshitz theory we show that the illumination of one
(Si) plate in the three-layer systems Au--ethanol--Si,
Si--ethanol--Si and $\alpha$-Al${}_2$O${}_3$--ethanol--Si
with laser pulses can change the Casimir attraction to Casimir
repulsion and vice versa. The proposed effect opens novel
opportunities in nanotechnology to actuate the periodic
movement in electro- and optomechanical micromachines based
entirely on the zero-point oscillations of the quantum vacuum
without the action of mechanical springs.
\end{abstract}
\pacs{12.20.-m, 12.20.Ds, 85.85.+j}

\section{Introduction}
Recent trends point towards an increased role of the Casimir effect in
both fundamental physics and nanotechnology. The Casimir force \cite{1}
acts between two closely spaced neutral metallic plates. It arises
due to the zero-point energy of the electromagnetic field. The Lifshitz
theory \cite{2} gave an accurate description of the material properties
in the Casimir force and represented it as the retarded limit
of the familiar van der Waals forces. According to current concepts,
Casimir forces act between metals, insulators and semiconductors,
between a molecule and a macrobody and between two molecules.
In the last two cases they are usually referred to as the
Casimir-Polder force. Multidisciplinary applications of the
Casimir force and the first precise measurements are reviewed in
\cite{3}. Later experiments [4--14]
stimulated
the development of new theoretical methods applicable to more
complicated configurations other than two static parallel plates 
[15--20].
The question whether the Casimir
force is always attractive or it can also be repulsive, as was
predicted for ideal metallic spherical and cubical shells
\cite{3}, has been debated \cite{14}.

The applications of the Casimir force are promising in the
design, fabrication and actuation of micro- and nanomechanical
devices. When the characteristic sizes of a device shrink below
a micrometer, the Casimir force may become larger than typical
electric forces. The first devices actuated by the Casimir force
were demonstrated in \cite{15,15a}. In \cite{16,16a} the lateral
Casimir force acting between corrugated surfaces was predicted.
Later it was demonstrated experimentally \cite{4,4a}. The lateral
Casimir force was recently used to propose the Casimir driven
ratchets and pinions \cite{17,18,18a}. Considerable opportunities
for micromechanical design would be opened by pulsating Casimir plates
moving back and forth entirely
due to the effect of the zero-point energy,
without the action of mechanical springs.
This can be
achieved only through use of both attractive and repulsive Casimir
forces. In this connection it should be noted that while the
repulsive Casimir forces for a single cube or a sphere are
still debated the Casimir repulsion between the
two parallel plates is well understood.
Repulsion occurs when the plates with
dielectric permittivities $\varepsilon_1$ and $\varepsilon_2$
along the imaginary frequency axis are immersed inside a
medium with dielectric permittivity $\varepsilon_0$ such that
$\varepsilon_1<\varepsilon_0<\varepsilon_2$ or
$\varepsilon_2<\varepsilon_0<\varepsilon_1$ \cite{19,20}.
At short separations in a nonretarded van der Waals regime
this effect was discussed for a long time and
measurements have been reported (see for example review \cite{20a}
and one of the later experiments \cite{20b}).

In this paper we consider three pairs of parallel plates
immersed in ethanol. First pair includes Au and Si plates with
the Si plate illuminated by light pulses from a laser.
The second pair consists of two similar Si plates with one
of them illuminated by light pulses.
As was recently shown in \cite{9}, the illumination with
light of appropriate power increases the charge carrier
density in Si by several orders of magnitude and
changes the dielectric permittivity over a wide frequency range
along the imaginary frequency axis leading
to the modulation of the Casimir force.
In the third pair, one plate is
made of $\alpha$-Al${}_2$O${}_3$ and the other of Si.
The latter plate is illuminated with laser
pulses. For all pairs of plates we calculate the Casimir
force per unit area as a function of separation distance.
It appears that within a wide range of separations there is
a repulsive Casimir force for the first pair of plates when
the light is off and an attractive force when the light is on.
For the second and third pairs of plates, the force is
repulsive when the light is on and attractive when the light
is off. Thus we find for the first time that illumination
with laser light can change Casimir attraction to Casimir
repulsion and vice versa. By appropriately choosing the
duration of the pulse and the time between pulses, it
is possible to obtain
pulsating Casimir plates.

\section{Theoretical approach and dielectric permittivities}
We consider the Casimir interaction between two plates with
dielectric permittivity $\varepsilon_1(\omega)$ and
$\varepsilon_2(\omega)$ immersed in ethanol with the
dielectric permittivity $\varepsilon_0(\omega)$ at a
temperature $T$ in thermal equilibrium. The separation
between the plates is $a$. The Casimir pressure is given
by the Lifshitz formula \cite{2,3}
\begin{eqnarray}
&&
P(a,T)=-\frac{k_BT}{8\pi a^3}
\sum\limits_{l=0}^{\infty}{\vphantom{\sum}}^{\prime}
\int_{\sqrt{\varepsilon_0}\zeta_l}^{\infty}y^2dy
\label{eq1} \\
&&
\times\left[
\frac{1}{e^yr_{\rm TM;1}^{-1}(\zeta_l,y)r_{\rm TM;2}^{-1}(\zeta_l,y)
-1}
+\frac{1}{e^yr_{\rm TE;1}^{-1}(\zeta_l,y)r_{\rm TE;2}^{-1}(\zeta_l,y)
-1}\right].
\nonumber
\end{eqnarray}
\noindent
Here the dimensionless Matsubara frequencies, $\zeta_l$, are related with
dimensional ones, $\xi_l$, by
$\zeta_l=2a\xi_l/c=4\pi k_BTal/(\hbar c)$, $l=0,\,1,\,2,\ldots\,$,
$k_B$ is the Boltzmann constant. The reflection coefficients
on the 1st and 2nd plates for TM and TE polarizations are
defined as
\begin{eqnarray}
&&
r_{\rm TM;1,2}(\zeta_l,y)=
\frac{\varepsilon_{1,2}y-\varepsilon_0\sqrt{y^2+(\varepsilon_{1,2}-
\varepsilon_0)\zeta_l^2}}{\varepsilon_{1,2}y+
\varepsilon_0\sqrt{y^2+(\varepsilon_{1,2}-\varepsilon_0)\zeta_l^2}},
\nonumber \\
&& \label{eq2} \\
&&
r_{\rm TE;1,2}(\zeta_l,y)=
\frac{\sqrt{y^2+(\varepsilon_{1,2}-
\varepsilon_0)\zeta_l^2}-y}{\sqrt{y^2+(\varepsilon_{1,2}-
\varepsilon_0)\zeta_l^2}+y},
\nonumber
\end{eqnarray}
\noindent
where $\varepsilon_{1,2}=\varepsilon_{1,2}(i\xi_l)$,
$\varepsilon_{0}=\varepsilon_{0}(i\xi_l)$.

To calculate the Casimir pressure for the above three pairs of
plates one needs the dielectric permittivities of Au, Si,
$\alpha$-Al${}_2$O${}_3$ and ethanol,
$\varepsilon^{Au}(i\xi)$, $\varepsilon^{Si}(i\xi)$,
$\varepsilon^{\alpha}(i\xi)$ and $\varepsilon_{0}(i\xi)$,
along the imaginary frequency axis. For Au and dielectric
Si (in the absence of laser light) the precise results for
permittivities are computed in \cite{21} by means of
the tabulated optical data for the complex index of
refraction \cite{22} and Kramers-Kronig relation. The obtained
permittivities are shown in Fig.~1 by the long-dashed line
and solid line 1, respectively. For ethanol and
$\alpha$-Al${}_2$O${}_3$ the dielectric permittivities along the
imaginary frequency axis can be presented in the
Ninham-Parsegian approximation \cite{19,20,22a}
\begin{equation}
\varepsilon^{(k)}(i\xi_l)=1+
\frac{C_k^{\rm IR}}{1+{\xi_l^2}/{\omega_{{\rm IR},k}^2}}+
\frac{C_k^{\rm UV}}{1+{\xi_l^2}/{\omega_{{\rm UV},k}^2}},
\label{eq3}
\end{equation}
where for ethanol
$C_1^{\rm IR}=23.84$, $C_1^{\rm UV}=0.852$,
$\omega_{{\rm IR},1}=6.600\times 10^{14}\,$rad/s,
$\omega_{{\rm UV},1}=1.140\times 10^{16}\,$rad/s ,
and for $\alpha$-Al${}_2$O${}_3$ it holds
$C_2^{\rm IR}=7.03$, $C_2^{\rm UV}=2.072$,
$\omega_{{\rm IR},2}=1.000\times 10^{14}\,$rad/s,
$\omega_{{\rm UV},2}=2.000\times 10^{16}\,$rad/s.
In Fig.~1 the dielectric permittivity of ethanol,
$\varepsilon^{(1)}(i\xi_l)\equiv\varepsilon_0(i\xi_l)$,
is shown by the short-dashed line, and the permittivity of
$\alpha$-Al${}_2$O${}_3$,
$\varepsilon^{(2)}(i\xi_l)\equiv\varepsilon^{\alpha}(i\xi_l)$,
is shown by the dotted line.

\begin{figure*}
\vspace*{-10cm}
\begin{center}
\hspace*{2cm}
\resizebox{1.\textwidth}{!}{%
\includegraphics{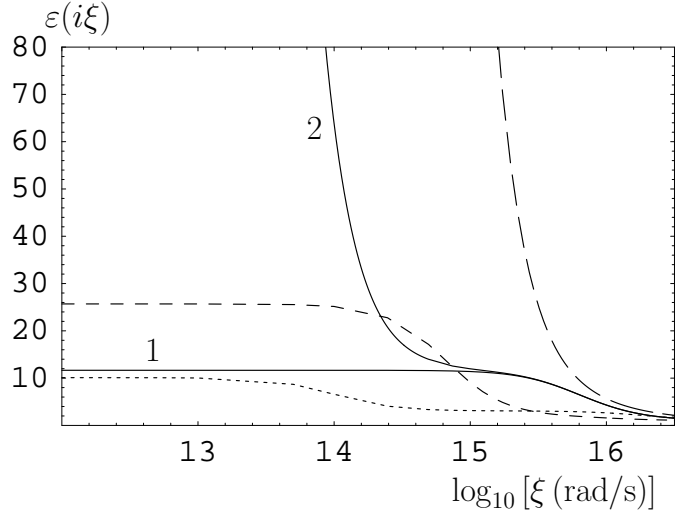}
}
\end{center}
\vspace*{-5.3cm}
\caption{The dielectric permittivities of different materials
along the imaginary frequency axis are shown with solid
lines 1 and 2 for Si in the absence and in the presence of laser
light, respectively, with a long-dashed line for Au, with a
short-dashed line for ethanol, and with a dotted line for
$\alpha$-Al${}_2$O${}_3$.
}
\end{figure*}
Careful attention should be paid to the influence of light
pulses on the Si plate. A few micron thick
single crystal membrane of $<100>$ orientation can be used as
a dielectric Si plate. The static dielectric permittivity
of such a plate is equal to
$\varepsilon^{Si}(0)\approx 11.66$ \cite{21,22} (see the
solid line 1 in Fig.~1). When the light from an Ar laser is
incident on the Si plate,
the equilibrium number of charge carriers per unit volume
is rapidly established, during a period of time much
shorter than the duration of the pulse \cite{9,23}.
The increase in the carrier density on illumination can lead
to electrostatic effects due to changes in the band bending
at the semiconductor surface. Therefore care must be taken
to achieve flat bands by surface passivation.

Assuming that there is an equilibrium concentration of
electrons and holes in the presence of light, we obtain
the dielectric permittivity
of Si in the form
\begin{equation}
\varepsilon_{L}^{Si}(i\xi_l)=\varepsilon^{Si}(i\xi_l)+
\frac{(\omega_p^{(e)})^2}{\xi_l\left(\xi_l+
\gamma^{(e)}\right)}+
\frac{(\omega_p^{(p)})^2}{\xi_l\left(\xi_l+
\gamma^{(p)}\right)}.
\label{eq4}
\end{equation}
\noindent
Here $\omega_p^{(e,p)}$ and $\gamma^{(e,p)}$ are the plasma
frequencies and relaxation parameters for electrons and holes,
respectively. The values of the relaxation parameters and the
effective masses of charge carriers are the following \cite{23}:
$\gamma^{(p)}\approx 5.0\times 10^{12}\,$rad/s,
 $\gamma^{(e)}\approx 1.8\times 10^{13}\,$rad/s,
$m_p^{\ast}=0.2063m_e$, $m_e^{\ast}=0.2588m_e$.
The values of the plasma frequencies can be found from the equation
$\omega_p^{(e,p)}=\left(
{n_Le^2}/{m_{e,p}^{\ast}\epsilon_0}\right)^{1/2}$,
where $\epsilon_0$ is the permittivity of vacuum and $n_L$ is
the density of charge carriers of each type in the presence
of light. The typical value of
$n_L\approx 2.1\times 10^{19}\,\mbox{cm}^{-3}$ was found
in \cite{9} for a light power of about 3.4\,mW
absorbed on a surface area $\pi w^2/4$, where $w=0.23\,$mm
is the diameter equal to the Gaussian width of the beam. Then
one obtains
$\omega_p^{(e)}\approx 5.08\times 10^{14}\,$rad/s and
$\omega_p^{(p)}\approx 5.69\times 10^{14}\,$rad/s.

\section{Computational results}
In Fig.~2 we present the computational results for the Casimir
pressure versus separation distance between the first pair
of plates, i.e., for Au and Si plates separated by ethanol,
where Si is illuminated with laser pulses.
These results are computed using Eq.~(\ref{eq1}) where
$\varepsilon_1(i\xi_l)=\varepsilon^{Au}(i\xi_l)$,
$\varepsilon_2(i\xi_l)=\varepsilon^{Si}(i\xi_l)$ in the
absence of laser light and
$\varepsilon_2(i\xi_l)=\varepsilon_L^{Si}(i\xi_l)$ in the
presence of laser light.
In both cases
$\varepsilon_0(i\xi_l)\equiv\varepsilon^{(1)}(i\xi_l)$
is the permittivity of ethanol.
The pressure-distance dependence in the absence of light on
a Si plate is shown as line 1. As is seen in Fig.~2, for
separations larger than 156\,nm the Casimir pressure shown
by line 1 is repulsive. This effect of repulsion in
a 3-layer system is well known \cite{19,20,20a,20b}.
It reflects the fact that in Fig.~1 the inequalities
$\varepsilon^{Si}(i\xi)<\varepsilon_0(i\xi)<\varepsilon^{Au}(i\xi)$
are valid within a wide frequency range. These three permittivities
are shown as the solid line 1, the short-dashed line and the
long-dashed line, respectively. In particular, from Eq.~(\ref{eq3})
it follows:
$\varepsilon_0(0)=1+C_1^{\rm IR}+C_1^{\rm UV}=25.692>
\varepsilon^{Si}(0)$.
The pressure-distance dependence in the presence of laser light
on a Si plate is shown by line 2 in Fig.~2. This line
corresponds to attraction at all separation distances.
The physical explanation of this fact can be obtained in Fig.~1
where the dielectric permittivity of Si in the presence of
light (the solid line 2)
$\varepsilon_L^{Si}(i\xi)>\varepsilon_0(i\xi)$ within a wide
frequency region. Thus, the illumination of a Si plate with
laser light changes the Casimir force from repulsion to
attraction.
\begin{figure*}
\vspace*{-11.7cm}
\begin{center}
\hspace*{2cm}
\resizebox{1.1\textwidth}{!}{%
\includegraphics{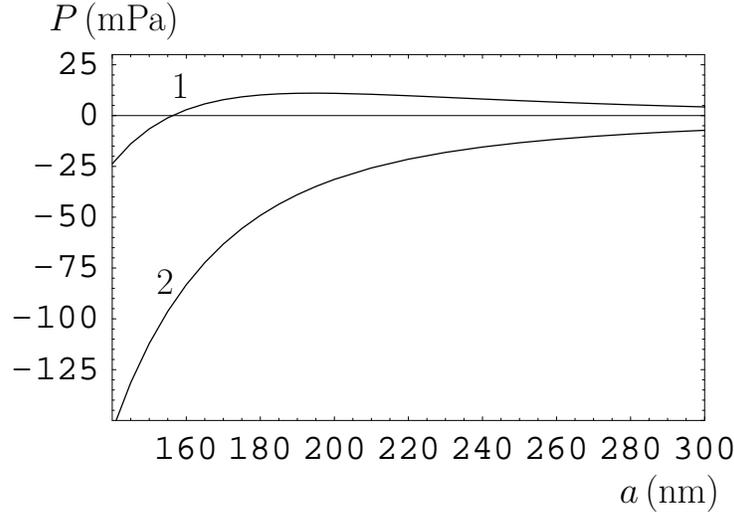}
}
\end{center}
\vspace*{-5.7cm}
\caption{The Casimir pressure versus separation in a three-layer
system Au--ethanol--Si with no light on the Si plate is shown by line 1
and with the illuminated Si plate is shown by line 2.
}
\end{figure*}
Note that the above result does not depend on discussions
in literature (see, e.g., \cite{24}) on the value of TE
reflection coefficient of metals at zero frequency.
The reason is that for insulators and semiconductors
$r_{\rm TE;2}(0,y)=0$ and thus the TE reflection coefficient of
Au, $r_{\rm TE;1}(0,y)$, regardless of its magnitude, does not
contribute to the Casimir pressure as it is multiplied
by zero.

Now we consider the computational results for the Casimir
pressure in the second pair of plates Si--ethanol--Si,
where one of the Si plates is illuminated with laser pulses.
In the absence of light
$\varepsilon_1(i\xi_l)=\varepsilon_2(i\xi_l)
=\varepsilon^{Si}(i\xi_l)$ where
$\varepsilon^{Si}$ is the permittivity of high resistivity Si.
The permittivity of ethanol is
$\varepsilon_0(i\xi_l)\equiv\varepsilon^{(1)}(i\xi_l)$
from Eq.~(\ref{eq3}).
Substituting this in Eq.~(\ref{eq1}) we arrive at the pressure-distance
relation shown in Fig.~3 by line 1. As is seen in Fig.~3, the
respective Casimir pressure is attractive at all separations. This
is expected from the outset because the permittivities of both
plates are equal. Now let the second Si plate be illuminated
with a laser pulse. Then in Eq.~(\ref{eq1}) the dielectric
permittivity
$\varepsilon_2(i\xi_l)=\varepsilon_L^{Si}(i\xi_l)$,
where $\varepsilon_L^{Si}$ is defined in Eq.~(\ref{eq4}).
In this case the computational results for the Casimir pressure
versus separation are shown by line 2 in Fig.~3. As is seen in the
figure, at $a<175\,$nm the Casimir force is attractive, but at
larger separations it becomes repulsive. The repulsion is explained
by the fact that within a wide frequency range the inequalities
$\varepsilon^{Si}(i\xi)<\varepsilon_0(i\xi)<\varepsilon_L^{Si}(i\xi)$
hold. Thus illumination with light leads to a change from
Casimir attraction to Casimir repulsion in the system of two
Si plates separated by ethanol.
\begin{figure*}
\vspace*{-11.5cm}
\begin{center}
\hspace*{2cm}
\resizebox{1.1\textwidth}{!}{%
\includegraphics{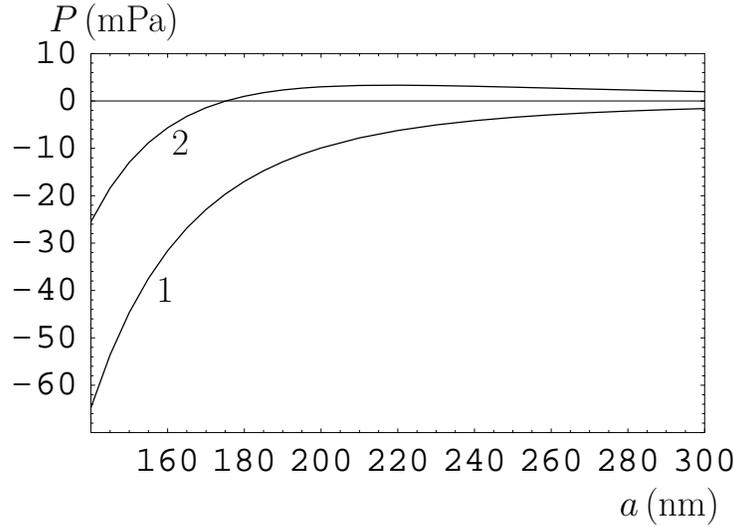}
}
\end{center}
\vspace*{-5.3cm}
\caption{The Casimir pressure versus separation in a three-layer
system Si--ethanol--Si with no light on both Si plates is shown by line 1
and with one illuminated Si plate is shown by line 2.
}
\end{figure*}

In the previously considered cases, the magnitudes of the
repulsive forces were several times less than the magnitude
of the attractive forces.
However, it is possible to design a case where the light-induced
Casimir repulsion is of the same order of magnitude as the
attraction. A good example is given by the three-layer system
$\alpha$-Al${}_2$O${}_3$--ethanol--Si, where the Si plate
is illuminated with laser pulses (the third pair of plates). 
First we perform the computations
of the Casimir pressure using Eq.~(\ref{eq1}) in the absence of
laser light. In this case
$\varepsilon_1(i\xi_l)=\varepsilon^{(2)}(i\xi_l)
\equiv\varepsilon^{\alpha}(i\xi_l)$, where the
dielectric permittivity $\varepsilon^{\alpha}$ of $\alpha$-Al${}_2$O${}_3$
is defined in Eq.~(\ref{eq3}),
$\varepsilon_2(i\xi_l)=\varepsilon^{Si}(i\xi_l)$, and
$\varepsilon_0(i\xi_l)\equiv\varepsilon^{(1)}(i\xi_l)$ from
Eq.~(\ref{eq3}) is the dielectric permittivity of ethanol.
The computational results for the Casimir pressure versus separation
distance are shown by the line 1 in Fig.~4.
It can be observed that the Casimir force is
attractive as expected because within a wide frequency range
both dielectric permittivities of Si, $\varepsilon^{Si}(i\xi)$,
 and of $\alpha$-Al${}_2$O${}_3$, $\varepsilon^{\alpha}(i\xi)$, are
smaller than the dielectric permittivity of ethanol
$\varepsilon_0(i\xi)$ (see Fig.~1).
For example, the static dielectric permittivity of
$\alpha$-Al${}_2$O${}_3$, as given by Eq.~(\ref{eq3}), is
$\varepsilon^{\alpha}(0)=1+C_2^{\rm IR}+C_2^{\rm UV}=10.102<
\varepsilon_0(0)$.

Now let the Si plate be illuminated with the laser pulse.
In this case we substitute into Eq.~(\ref{eq1})
$\varepsilon_2(i\xi_l)\equiv\varepsilon_L^{Si}(i\xi_l)$,
as defined in Eq.~(\ref{eq4}), and with the other two permittivities
kept unchanged. The computational results for the Casimir pressure
versus separation distance are shown by the line 2 in Fig.~4.
As is seen from the figure, at separations $a>71.5\,$nm
the corresponding Casimir force is repulsive. Remarkably, in this
case the Casimir repulsion and attraction are of the same
order of magnitude within a wide range of separations.
Thus the third pair of plates provides an example where the
illumination of the Si plate changes the Casimir attraction to
a repulsive force of the same order.
For the
observation of the pulsating Casimir force we envision
that the plates would be
 completely immersed in the liquid far away from any
air-liquid interfaces, thus, preventing the occurrence
of capillary forces. Surface
preparation of the plates will be necessary to bring
about intimate contact between the plates and the
liquid. The only liquid based force is the
drag force due to the movement of the plates in
response to the change in the force. For pressure
values of around 10 mPa and typical spring constants
of 0.02 N/m, the corresponding drag pressure from
plate movement would be 6 orders of magnitude less in
value.
\begin{figure*}
\vspace*{-11.5cm}
\begin{center}
\hspace*{2cm}
\resizebox{1.1\textwidth}{!}{%
\includegraphics{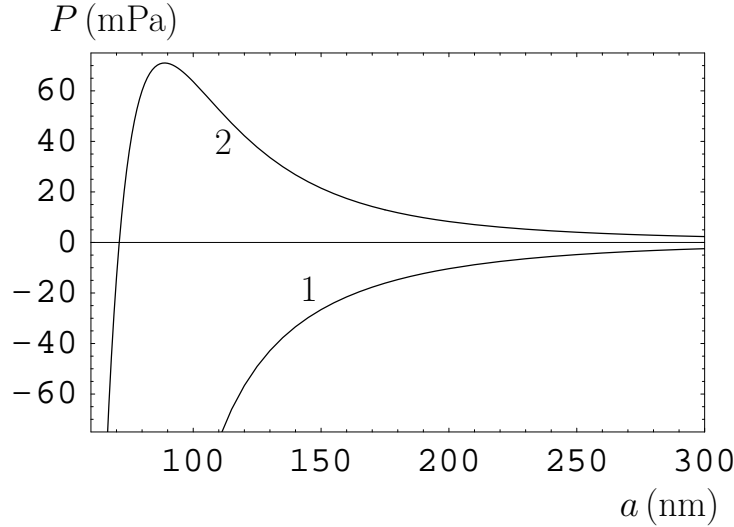}
}\end{center}
\vspace*{-5.7cm}
\caption{The Casimir pressure versus separation in a three-layer
system $\alpha$-Al${}_2$O${}_3$--ethanol--Si with no light on
the Si plate is shown by line 1 and with the illuminated Si plate
is shown by line 2.
}
\end{figure*}

\section{Conclusions and discussion}
In the above we have shown that the illumination of one of the
plates in a three-layer Casimir system can change the repulsion
to attraction and vice versa entirely due to the modification
of the spectrum of the quantum vacuum.
This is attained by the
combination of the familiar properties of three-layer
systems \cite{19} and the recently demonstrated modulation
of the Casimir force with laser light \cite{9}.
The proposed effect of the pulsating
Casimir force can be used to actuate the periodic movement
of electrodes and mirrors in electro- and optomechanical
micromachines. This can be achieved by using the standard
frequency generators and modulators to select the appropriate
duration and time between the laser pulses.
For the case of ethanol
considered here the hydrodynamic and viscous effects
will limit the operational bandwidth of the actuators
to around 10 MHz for a plate size of 10x10 microns.
Here the deviations from plate parallelism of less
than 0.2 degrees (which is experimentally achievable),
would lead to a less than 3\% change in
the value of the Casimir pressure computed above \cite{3}
with no change in its sign.
The pulsating Casimir force will also find applications
in nanotechnology for controlling the efficiency and increasing
the operational bandwidth of microswitches, micromirrors
and nanotweezers.

\section*{Acknowledgments}
The work on the illumination induced
transitions between the attractive and repulsive Casimir
regimes  was supported by the
DOE Grant No.~DE-FG02-04ER46131.
Numerical computations on the Casimir pressure were
supported by the NSF Grant No.~PHY0653657.
G.L.K. and V.M.M. were also partially supported by
DFG Grant No.~436\,RUS\,113/789/0--3.


\section*{References}
\numrefs{99}
\bibitem {1}
Casimir H B G 1948
{\it Proc. K. Ned. Akad. Wet.}
{\bf 51} 793 
\bibitem{2}
Lifshitz E M 1956
{\it Sov. Phys. JETP}  {\bf 2} 73
\bibitem{3}
Bordag M, Mohideen U, and Mostepanenko V M 2001
{\it Phys. Rep.} {\bf 353} 1 
\bibitem{4}
Chen F, Mohideen U, Klimchitskaya G L, and
Mos\-te\-pa\-nen\-ko V M 2002
{\it Phys. Rev. Lett.} {\bf 88} 101801
\bibitem{4a}
Chen F, Mohideen U, Klimchitskaya G L, and
Mos\-te\-pa\-nen\-ko V M 2002
{\it Phys. Rev.} A {\bf 66} 032113 
\bibitem{5}
Bressi G, Carugno G, Onofrio R, and Ruoso G 2002
{\it Phys. Rev. Lett.} {\bf 88} 041804 
\bibitem{6}
Decca R S, Fischbach E, Klimchitskaya G L,
 Krause D E, L\'opez D, and Mostepanenko V M 2003
{\it Phys. Rev.} D {\bf 68}, 116003 
\bibitem{6a}
Decca R S, L\'opez D, Fischbach E, Klimchitskaya G L,
 Krause D E and Mostepanenko V M 2005
 {\it  Ann. Phys. NY } {\bf 318} 37
\bibitem{6b}
Decca R S, L\'opez D, Fischbach E, Klimchitskaya G L,
 Krause D E and Mostepanenko V M 2007
 {\it  Phys. Rev} D {\bf 75} 077101 
\bibitem{6c}
Decca R S, L\'opez D, Fischbach E, Klimchitskaya G L,
 Krause D E and Mostepanenko V M 2007
arXiv:0706.3283; {\it Eur. Phys. J} C, to appear
\bibitem{7}
Chen F, Mohideen U, Klimchitskaya G L, and
Mos\-te\-pa\-nen\-ko V M 2005
{\it Phys. Rev.} A {\bf 72} 020101(R) 
\bibitem{7a}
Chen F, Mohideen U, Klimchitskaya G L, and
Mos\-te\-pa\-nen\-ko V M 2005
{\it Phys. Rev.} A
{\bf 74} 022103 
\bibitem{8}
Chen F,  Klimchitskaya G L,
Mos\-te\-pa\-nen\-ko V M, and Mohideen U 2006
{\it Phys. Rev. Lett.}  {\bf 97} 170402 
\bibitem{9}
Chen F,  Klimchitskaya G L,
Mos\-te\-pa\-nen\-ko V M, and Mohideen U 2007
{\it Optics Express} {\bf 15} 4823 
\bibitem{10}
Emig T, Jaffe R L, Kardar M, and Scardicchio A 2006
{\it Phys. Rev. Lett.} {\bf 96} 080403
\bibitem{11}
Bulgac A, Magierski P, and Wirzba A 2006
{\it Phys. Rev.} D {\bf 73} 025007 
\bibitem{12}
Bordag M 2006 {\it Phys. Rev. D} {\bf 73} 125018
\bibitem{13}
Gies H and Klingm\"{u}ller K 2006
{\it Phys. Rev. Lett.} {\bf 96} 220401
\bibitem{13a}
Gies H and Klingm\"{u}ller K 2006
{\it Phys. Rev.} D {\bf 74}, 045002
\bibitem{13b}
Haro J and Elizalde E 2006
{\it Phys. Rev. Lett.} {\bf 97} 130401
\bibitem{14}
Hertzberg M P, Jaffe R L, Kardar M, and Scardicchio A 2005
{\it Phys. Rev. Lett.} {\bf 95} 250402 
\bibitem{15}
Chan H B, Aksyuk V A, Kleiman R N, Bishop D J, and
Capasso F 2001
{\it Science} {\bf 291} 1941 
\bibitem{15a}
Chan H B, Aksyuk V A, Kleiman R N, Bishop D J, and
Capasso F 2001
{\it Phys. Rev. Lett.} {\bf 87} 211801 
\bibitem{16}
Golestanian R and Kardar M 1997
{\it Phys. Rev. Lett.} {\bf 78} 3421 
\bibitem{16a}
Golestanian R and Kardar M 1998
{\it Phys. Rev.} A {\bf 58}, 1713 (1998).
\bibitem{17}
Emig T 2007
{\it Phys. Rev. Lett.} {\bf 98} 160801 
\bibitem{18}
Ashourvan  A, Miri M, and Golestanian R 2007
{\it Phys. Rev. Lett.} {\bf 98} 140801 
\bibitem{18a}
Ashourvan  A, Miri M, and Golestanian R 2007
{\it Phys. Rev.} E {\bf 75} 040103(R) 
\bibitem{19}
Mahanty J and Ninham B W 1976
{\it Dispersion Forces} (New York: Academic)
\bibitem{20}
Munday J N, Iannuzzi D, Barash Y, and Capasso F 2005
{\it Phys. Rev.} A {\bf 71} 042102
\bibitem{20a}
Visser J 1981
{\it Adv. Coll. Interface Sci.} {\bf 15} 157 
\bibitem{20b}
Meurk A, Luckham P F, and Bergstr\"{o}m L 1997
{\it Langmuir} {\bf 13} 3896
\bibitem{21}
Caride A O, Klimchitskaya G L, Mostepanenko V M,
and Zanette S I 2005
{\it Phys. Rev.} A {\bf 71} 042901 
\bibitem {22}
Palik E D (ed.) 1985 {\it Handbook of Optical Constants of Solids}
(New York: Academic Press)
\bibitem {22a}
Bergstr\"{o}m L 1997
{\it Adv. Coll. Interface Sci.} {\bf 70} 125 
\bibitem{23}
Vogel T, Dobel G, Holzhauer E, Salzmann H, and Theurer A 1992
{\it Appl. Opt.} {\bf 31} 329 
\bibitem{24}
Bezerra V B, Decca R S, Fischbach E, Geyer B,
Klimchitskaya G L, Krause D E, L\'opez D,
Mostepanenko V M and Romero C 2006
{\it Phys. Rev.} E {\bf 73} 0281101
\endnumrefs
\end{document}